\documentclass[superscriptaddress,preprint]{revtex4-2}

\usepackage{mathpazo}
\usepackage[T1]{fontenc}
\usepackage[latin9]{inputenc}
\usepackage{color}
\usepackage{multirow}
\usepackage{amsmath}
\usepackage{amssymb}
\usepackage{graphicx}
\usepackage{esint}
\usepackage[unicode=true,pdfusetitle,
 bookmarks=false,
 breaklinks=true,pdfborder={0 0 0},pdfborderstyle={},backref=false,colorlinks=true,pdfpagemode=FullScreen]
 {hyperref}
\hypersetup{
 colorlinks,citecolor=blue,filecolor=black,linkcolor=red,urlcolor=blue}

\makeatletter

\providecommand{\tabularnewline}{\\}

\usepackage{booktabs}
\usepackage{color}
\usepackage{units}
\usepackage{multirow}
\usepackage{makecell}
\usepackage{caption}

\usepackage{array}
\usepackage{breakurl}
\usepackage{color}

\makeatother

\begin{document}

\title{Spontaneous symmetry breaking and vortices in a tri-core nonlinear
fractional waveguide}

\author{Mateus C. P. dos Santos}
\email{mateuscalixtopereira@gmail}
\address{Instituto de Física, Universidade Federal de Goiás, 74.690-900, Goiânia,
Goiás, Brazil}

\author{Wesley B. Cardoso}
\address{Instituto de Física, Universidade Federal de Goiás, 74.690-900, Goiânia,
Goiás, Brazil}

\author{Dmitry V. Strunin}
\address{School of Mathematics, Physics and Computing, University of Southern
Queensland, Toowoomba, Queensland 4350, Australia}

\author{Boris A. Malomed}
\address{Department of Physical Electronics, School of Electrical Engineering,
Faculty of Engineering, and the Center for Light-Matter Interaction,
Tel Aviv University, P.O.B. 39040, Ramat Aviv, Tel Aviv, Israel}
\address{Instituto de Alta Investigación, Universidad de Tarapacá, Casilla
7D, Arica, Chile}

\begin{abstract}
We introduce a waveguiding system composed of three linearly-coupled
fractional waveguides, with a triangular (prismatic) transverse structure.
It may be realized as a tri-core nonlinear optical fiber with fractional
group-velocity dispersion (GVD), or, possibly, as a system of coupled
Gross--Pitaevskii equations for a set of three tunnel-coupled cigar-shaped
traps filled by a Bose-Einstein condensate of particles moving by
Lévy flights. The analysis is focused on the phenomenon of spontaneous
symmetry breaking (SSB) between components of triple solitons, and
the formation and stability of vortex modes. In the self-focusing
regime, we identify symmetric and asymmetric soliton states, whose
structure and stability are determined by the Lévy index of the fractional
GVD, the inter-core coupling strength, and the total energy, which
determines the system's nonlinearity. Bifurcation diagrams (of the
supercritical type) reveal regions where SSB occurs, identifying the
respective symmetric and asymmetric ground-state soliton modes. In
agreement with the general principle of the SSB theory, the solitons
with broken inter-component symmetry prevail with the increase of
the energy in the weakly-coupled system. Three-components vortex solitons
(which do not feature SSB) are studied too. Because the fractional
GVD breaks the system's Galilean invariance, we also address mobility
of the vortex solitons, by applying a boost to them. 
\end{abstract}

\maketitle

\section{Introduction}

Nonlinear Schrödinger (NLS) equations are a set of universal models
governing the wave propagation in dispersive nonlinear media \citep{ZMNP,Dauxois},
such as optical fibers \citep{Agrawal_13}, bulk media and photonic
crystals \citep{Kivshar_03}, Langmuir waves in plasmas \citep{Zakharov_JETP72,Ichikawa},
matter waves Bose-Einstein condensates (BECs) \citep{Pitaevskii_03,Pethick_08},
magnetics \citep{Laksh}, surface waves in fluids \citep{surface-waves,surface-waves2}
and solids \citep{Maugin}, etc. Commonly known solutions of the NLS
equations are fundamental solitons and higher-order ones (breathers
\citep{SY}), which have been created experimentally in a great variety
of physical setups.

A natural extension of the single NLS equation is a system of linearly
coupled ones, which describe copropagation of nonlinear waves in tunnel-coupled
channels, a well-known example being dual-core \citep{Jensen,Maier,Trillo,Peng}
and tri-core \citep{Akhmediev_JOS94,Arik} optical fibers. Recently,
much interest was drawn to nonlinear optics in multi-core fibers,
that support various multi-mode propagation regimes \citep{Agrawal,ultrahigh,Wabnitz,Wise}.

Nonlinear dual-core systems support obvious states in the form of
two-component solitons which are symmetric with respect to the coupled
cores. An effect induced by the intra-core self-focusing nonlinearity
in dual-core systems is the spontaneous symmetry breaking (SSB). It
occurs above a critical value of the soliton's energy, when the symmetric
solitons with equal components lose their stability and are replaced
by asymmetric ones with unequal components. This effect was studied
in detail theoretically \citep{Trillo,Peng}, and recently demonstrated
experimentally in dual-core optical fibers \citep{Ignac}.

A somewhat similar realization of soliton SSB was studied in models
based on a single NLS equation with the cubic self-focusing nonlinearity
concentrated at two mutually symmetric points in the form of delta-functions
\ \citep{Dong,Shatrughna}, as well as in their two-component version
\citep{Acus}. In those models, SSB was realized as spontaneous establishment
of a stationary structure in the form of two mutually asymmetric spikes
with unequal amplitudes, pinned to the two delta-functions.

Another novel direction for the study of NLS-like solitons is focused
on the fractional NLS (FNLS) equations. They were originally derived,
in the linear form, by dint of the Feynman-integral formalism as quantum-mechanical
equations for particles moving by Lévy flights in the classical limit
\citep{Laskin_PLA00,Laskin_PRA02,GuoXu,Laskin_18}. While such a quantum-mechnical
setting has not been yet realized experimentally, it was later proposed
to implement essentially the same linear equations as classical ones
for the propagation of optical beams in the paraxial approximation,
emulating the effect of the fractional diffraction by a phase shift
introduced for different spectral components by means of a properly
designed phase plate \citep{Longhi_OL15}. The optical realization
of the fractional linear Schrödinger equation in the temporal domain,
i.e., in a fiber cavity, has been reported in Ref. \citep{Shilong},
where the phase shifts emulating the action of the fractional group-velocity
dispersion (FGVD) was created as a computer-generated hologram.

The possibility to implement the fractional diffraction/dispersion in optics suggests one to include the nonlinearity of the dielectric material, the respective model being naturally based on
FNLS equations. This possibility has been elaborated theoretically
in great detail. The predicted effects include modulational instability
of continuous waves \citep{Zhang_CNSNS17}, many varieties of quasi-linear
modes \citep{Zhong_PRE16,Zhong_AP16} and solitons \citep{Secchi_AA14,Zhong_PRE16,Zhong_AP16,Hong_N17,Chen_PRE18,Wang_EPL18,HS},
including gap solitons in optical lattices \citep{Huang_OL16,Xiao_OE18},
multipole and multipeak modes \citep{Zeng_OL19,Qiu_CSF20,Li_OE20,Zeng_CSF21,Santos_CJP24},
soliton clusters \citep{Zeng_CP20,Wang_JO20}{, PT-symmetric solitons
\citep{Zhong_PD23,Zhong_CHAOS23}}, and solitary vortices \citep{Li_CSF20,Wang_JO20}.
Reviews of the theoretical results for solitons in models based on
FNLS and related equations are offered by Refs. \citep{review} and
\citep{review2}.

The SSB effect for two-component solitons in fractional dual-core
couplers has also been addressed \citep{Zeng_CSF20,Zeng_ND21,Strunin}.
The next natural step is to consider a tri-core linearly-coupled system
with the combination of FGVD and cubic self-focusing acting in each
core, which is modeled by a system of three coupled FNLS equations
(note that the experimental realization of FGVD, reported in Ref.
\citep{Shilong} for single-core fiber cavities, can be readily implemented
for multi-core systems as well). This is the subject of the present
work. The system of coupled equations is introduced in Section \ref{Sec2},
which is followed by the analysis of the SSB phenomenology and families
of three-component solitons in this system, reported in Section \ref{Sec3}.
Vortex solitons, with winding number $1$ carried by the triangular
(prismatic) set of three complex components (cf. Ref. \citep{Sigler}),
are considered in Section \ref{ivort}, where their stability area
is identified, and it is found that the vortex solitons do not give
rise to SSB, i.e., they are extremely robust modes. In the same section,
mobility of vortex solitons is addressed too, by means of systematic
simulations. The paper is concluded by Section \ref{Sec4}.

\section{The model \label{Sec2}}

\begin{figure}[tb]
\centering \includegraphics[width=0.8\columnwidth]{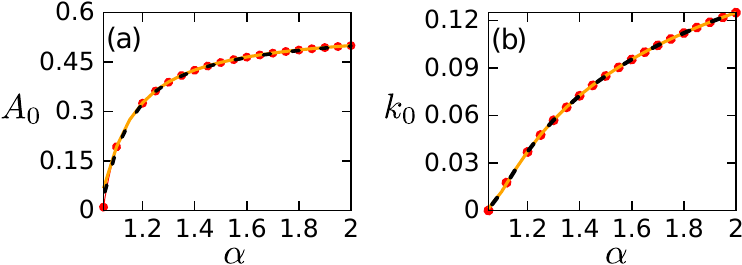} 
\caption{Coefficients $A_{0}$ (a) and $k_{0}$ (b) from Eqs. (\ref{A2})
and (\ref{k}) versus $\alpha$. The results, as obtained
from numerically found ground-states (GS) solutions of Eq. (\ref{single})
for different energies, $E=1$, $1.3$ and $1.5$, are displayed by
solid lines (orange), dotted lines (black) and circles (red), respectively.
The identical equality of these coefficients for all values of $E$
corroborates the validity of scaling relations (\ref{A2})
and (\ref{k}).}
\label{F0} 
\end{figure}

\begin{figure*}[!h]
\centering \includegraphics[width=0.82\textwidth]{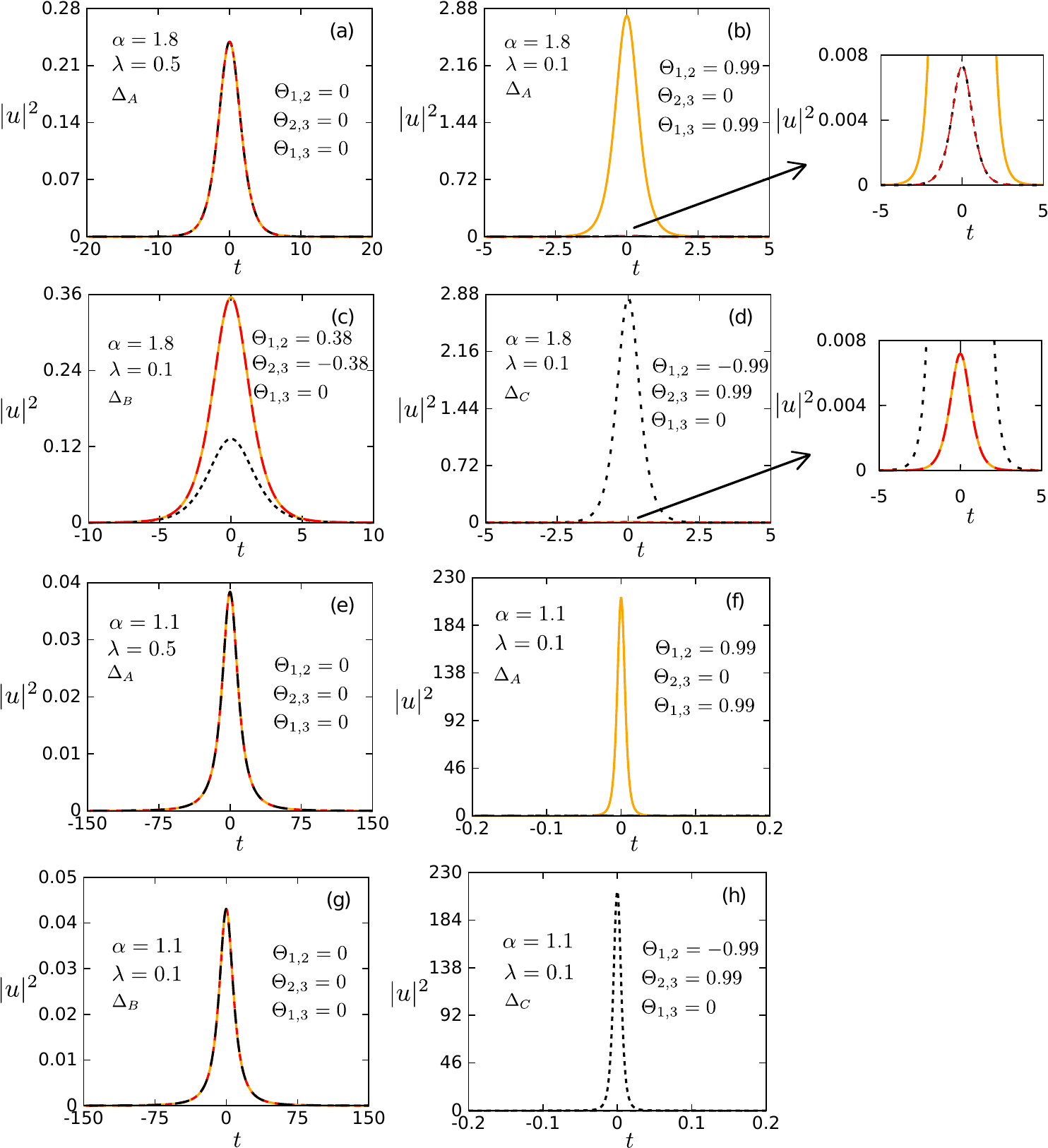} 
\caption{Symmetric and asymmetric GS (ground-state) solitons in the three-core
system with total energy $E=3$ are shown by means of their power
profiles, $|u_{1}(t)|^{2}$, $\left\vert u_{2}(t)\right\vert ^{2}$,
and $\left\vert u_{3}(t)\right\vert ^{2},$ which are plotted by orange
solid lines, black dotted lines, and red dashed lines, respectively.
They were obtained by the imaginary-time simulations of Eq. (\ref{EQ1})
from the inputs (\ref{AS1}), (\ref{AS2}), and
(\ref{AS3}), as indicated by symbols $\Delta_{A,B,C}$ in
each panel (values of LI $\alpha$ and linear-coupling coefficient
$\lambda$ are also indicated in the panels). The solitons in panels
(a, b) and (d-h) are stable, while the one in panel (c) is unstable.
The asymmetric solitons in panels (b) and (d) exhibit bistability,
belonging to the coexistence region shown in Fig. \ref{FBs1}(a).
The solitons in panels (f-h) realize a tristability case, belonging
to the highlighted region of Fig. \ref{FBs2}(a).}
\label{F1} 
\end{figure*}

\subsection{Three linearity coupled fractional nonlinear Schrödinger (FNLS) equations}

We consider the model of the tri-core optical fiber with the triangular
(prismatic) cross-section structure and amplitudes $U_{1,2,3}\left(t,z\right)$
of the optical waves propagating in the three linearly-coupled cores
with the FGVD and Kerr self-focusing nonlinearity carried by each
core. In the scaled form, the corresponding system of the coupled
FNLS equations takes the form (cf. similar systems for the tri-core
fibers with the regular (non-fractional) dispersion \citep{Longhi_OL15,Akhmediev_JOS94}:

\begin{align}
i\frac{\partial U_{1}}{\partial z}=\frac{1}{2}\left(-\frac{\partial^{2}}{\partial t^{2}}\right)^{\alpha/2}U_{1}-|U_{1}|^{2}U_{1}-\lambda\left(U_{2}+U_{3}\right),\nonumber \\
i\frac{\partial U_{2}}{\partial z}=\frac{1}{2}\left(-\frac{\partial^{2}}{\partial t^{2}}\right)^{\alpha/2}U_{2}-|U_{2}|^{2}U_{2}-\lambda\left(U_{3}+U_{1}\right),\nonumber \\
i\frac{\partial U_{3}}{\partial z}=\frac{1}{2}\left(-\frac{\partial^{2}}{\partial t^{2}}\right)^{\alpha/2}U_{3}-|U_{3}|^{2}U_{3}-\lambda\left(U_{1}+U_{2}\right).\nonumber \\
\label{EQ1}
\end{align}
Here the evolutional variable $z$ is, as usual \citep{Agrawal_13}
the propagation distance, $t$ is the temporal coordinate, and $\lambda>0$
is the coupling constant. By means of rescaling, the Kerr-nonlinearity
and FGVD coefficients in Eqs. (\ref{EQ1}) are fixed to be $+1$ or
$-1$ (the signs of the coefficients imply cubic self-focusing and
anomalous dispersion~\citep{Agrawal_13}). FGVD is represented by
the fractional Riesz derivative \citep{Riesz}, which is a pseudo-differential
(actually, integral) operator, defined as the juxtaposition of the
direct and inverse Fourier transforms, 
\begin{equation}
\left(-\partial^{2}/\partial t^{2}\right)^{\alpha/2}U=\frac{1}{2\pi}\int_{-\infty}^{+\infty}d\omega|\omega|^{\alpha}\int_{-\infty}^{+\infty}d\tau e^{-i\omega(\tau-t)}U(\tau),\label{Riesz-derivative}
\end{equation}
with the Lévy index (LI) $\alpha$ \citep{Mandelbrot}, which takes
values $1<\alpha\leq2$. With regard to the definition (\ref{Riesz-derivative}),
the Hamiltonian of system (\ref{EQ1}) can be written as
\begin{gather}
H=\sum_{j=1,2,3}\left[\frac{1}{2\pi}\int_{0}^{\infty}\omega^{\alpha}d\omega\int_{-\infty}^{+\infty}dt\int_{-\infty}^{+\infty}d\tau\cos\left(\omega(t-\tau)\right)\right.\nonumber \\
\left.\times U_{j}^{\ast}(\tau)U_{j}(t)-\frac{1}{2}\int_{-\infty}^{+\infty}dt\left\vert U_{j}(t)\right\vert ^{4}\right]\nonumber \\
-\lambda\int_{-\infty}^{+\infty}dt\sum_{j\neq k}U_{j}^{\ast}(t)U_{k}(t),\label{H}
\end{gather}
where $\ast$ stands for the complex conjugate.

\begin{figure*}[t]
\centering \includegraphics[width=0.9\textwidth]{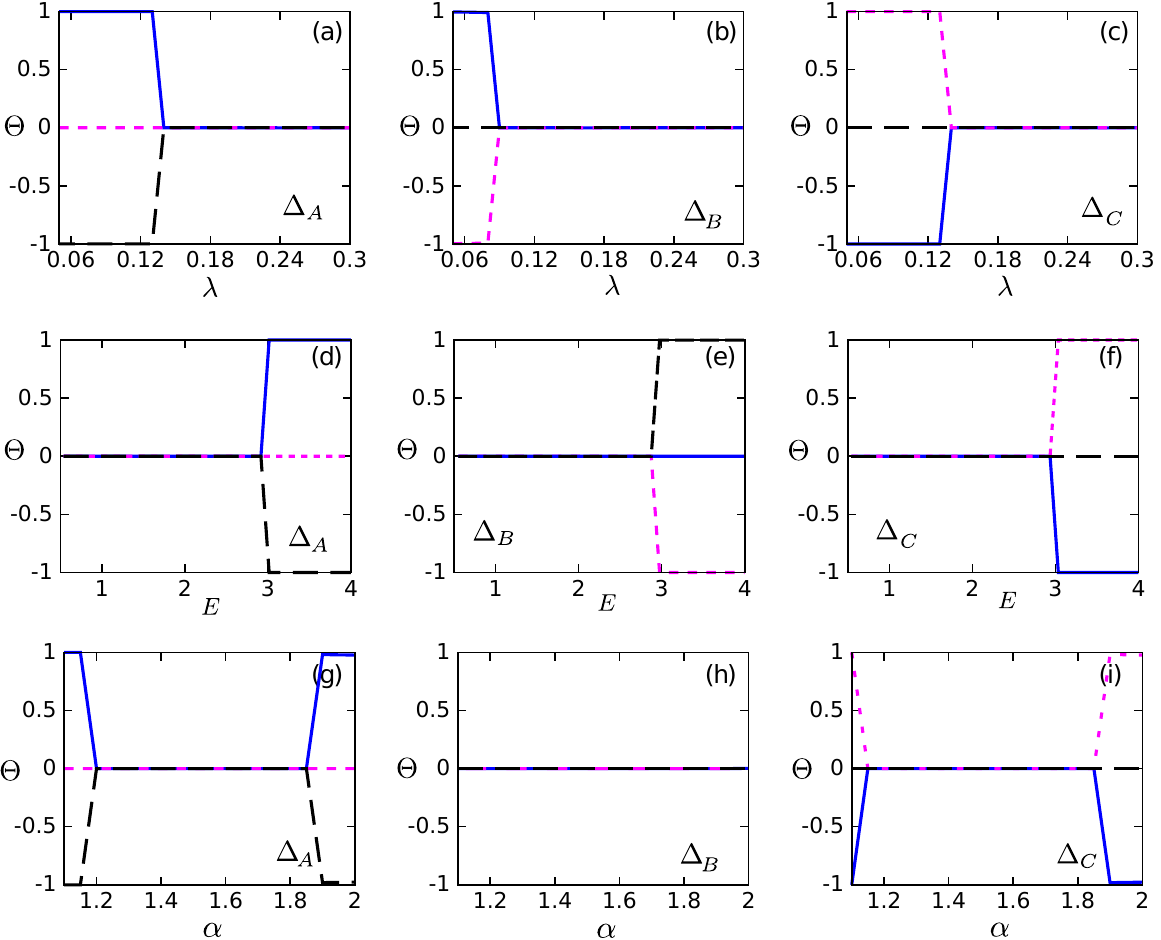} \caption{Asymmetry ratios (\ref{Theta}) $\Theta$ plotted versus
$\lambda$, $E$ and $\alpha$, for the GS solitons by the different
inputs of types $\Delta_{A}$, $\Delta_{B}$ and $\Delta_{C}$ (see
Eq. (\ref{AS1}) - (\ref{AS3})). The results for
$\Theta_{1,2}$, $\Theta_{2,3}$ and $\Theta_{3,1}$ are displayed
by solid (blue), dotted (magenta) and dashed (black) lines, respectively.
The other parameters are: (a-c) $\alpha=1.1$ and $E=3$; (d-f) $\alpha=1.1$
and $\lambda=0.13$; and (g-i) $E=3$ and $\lambda=0.13$.}
\label{F3} 
\end{figure*}

The classical (non-fractional) derivative corresponds to $\alpha=2$.
Smaller values, $\alpha\leq1$ give rise to the wave collapse in the
FNLS equation, which makes all solitons unstable \citep{Chen_PRE18}).

One may conjecture that the same system (\ref{EQ1}), with $z$ replaced
by scaled time, and $t$ replaced by a scaled coordinate, may appear
as Gross-Pitaevskii equations for mean-field wave functions of the
BEC of quantum particles that are governed, at the individual level,
by the fractional linear Schrödinger equation, in the case when the
BEC is loaded into a tunnel-coupled set of parallel cigar-shaped potential
traps that form a triangular (prismatic) structure, cf. Ref. \citep{Hidetsugu}.
However, a consistent derivation of such fractional Gross-Pitaevskii
equations was not reported, as yet.

Non-topological (zero-vorticity) soliton solutions to Eq. (\ref{EQ1})
with real propagation constant $k$ are looked for as
\begin{equation}
U_{1,2,3}\left(t,z\right)=e^{ikz}u_{1,2,3}\left(t\right),\label{U123}
\end{equation}
where $u_{1,2,3}\left(t\right)$ must be real localized solutions
of the system of three coupled equations:
\begin{align}
ku_{1}+\frac{1}{2}\left(-\frac{d^{2}}{dt^{2}}\right)^{\alpha/2}u_{1}-u_{1}^{3}-\lambda\left(u_{2}+u_{3}\right)=0,\nonumber \\
ku_{2}+\frac{1}{2}\left(-\frac{d^{2}}{dt^{2}}\right)^{\alpha/2}u_{2}-u_{2}^{3}-\lambda\left(u_{3}+u_{1}\right)=0,\nonumber \\
ku_{3}+\frac{1}{2}\left(-\frac{d^{2}}{dt^{2}}\right)^{\alpha/2}u_{3}-u_{3}^{3}-\lambda\left(u_{1}+u_{2}\right)=0.\label{realU}
\end{align}
It is expected that the system (\ref{realU}) admits more than a single
species of asymmetric solitons, as a result of SSB, cf. Ref. \citep{Arik},
where a similar problem was considered for a triangular set of linearly-coupled
Bragg gratings (with the usual first-order derivatives, instead of
the second-order ones or their fractional counterparts).

\begin{figure*}[t]
\centering \includegraphics[width=0.8\textwidth]{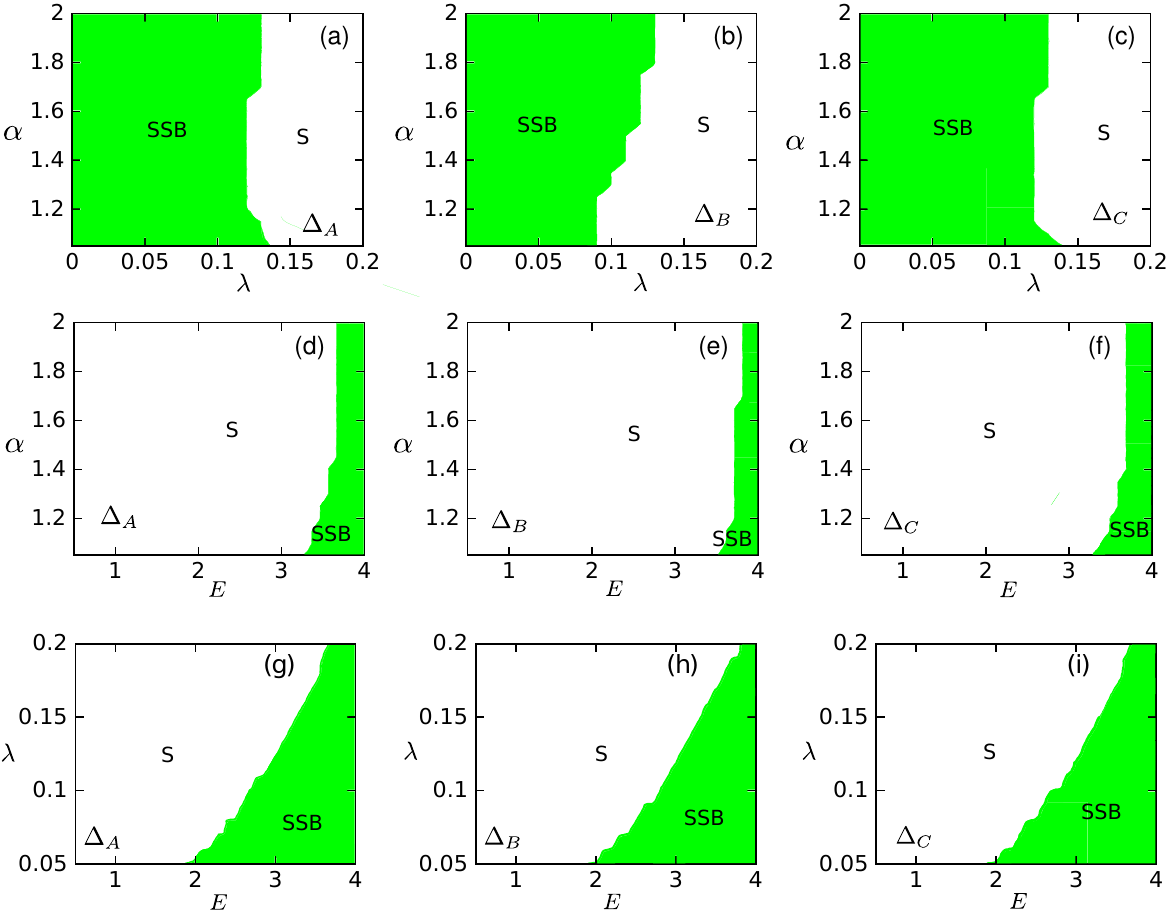} \caption{Phase diagrams for (a)symmetric solitons produced by inputs of types
$\Delta_{A}$, $\Delta_{B}$ and $\Delta_{C}$, see Eqs. (\ref{AS1}),
(\ref{AS2}), and (\ref{AS3}). The S (white) and
SSB (green) areas are populated by stable symmetric and asymmetric
GS solitons, respectively. The parameters used here are: (a-c) $E=3$;
(d-f) $\lambda=0.2$, and (g-i) $\alpha=1.7$.}
\label{F4} 
\end{figure*}

Soliton solutions of Eq. (\ref{realU}) are characterized by their
total energy (norm), 
\begin{equation}
E=\sum_{j=1,2,3}E_{j}\equiv\sum_{j=1,2,3}\int_{-\infty}^{+\infty}|U_{j}|^{2}dt,\label{Energy}
\end{equation}
where $E_{j}$ is the energy carried by each core. In the present
conservative system, $E$ is a dynamically invariant of Eq. (\ref{EQ1}),
while partial energies $E_{j}$ are not, as the cores can exchange
the energy through the linear coupling.

The structure of solutions with unequal components $u_{j}$ produced
by system (\ref{realU}) is characterized by the set of \textit{asymmetry
ratios,} 
\begin{equation}
\Theta_{n,m}=\frac{E_{n}-E_{m}}{E_{n}+E_{m}},\label{Theta}
\end{equation}
which are defined as relative difference in the energies of the optical
fields in different cores. These SSB parameters take values $-1\leq\Theta_{n,m}\leq+1$.
Symmetric solitons with equal energies in all cores have, obviously,
$\Theta_{n,m}=0$ for all pairs of indices $\left(m,n\right)$. On
the other hand, limit values $\Theta_{n,m}=1$ or $\Theta_{n,m}=-1$
indicate that the core $m$ or $n$ is empty, carrying no energy.

{Thus, while the asymmetry parameter of the usual double-core couplers
is a scalar, it is a three-component vector in the case of the three-core
system, which makes the results quite different, as shown below.}

\subsection{Scaling for single-component fractional solitons}

\begin{figure*}[t]
\centering \includegraphics[width=0.8\textwidth]{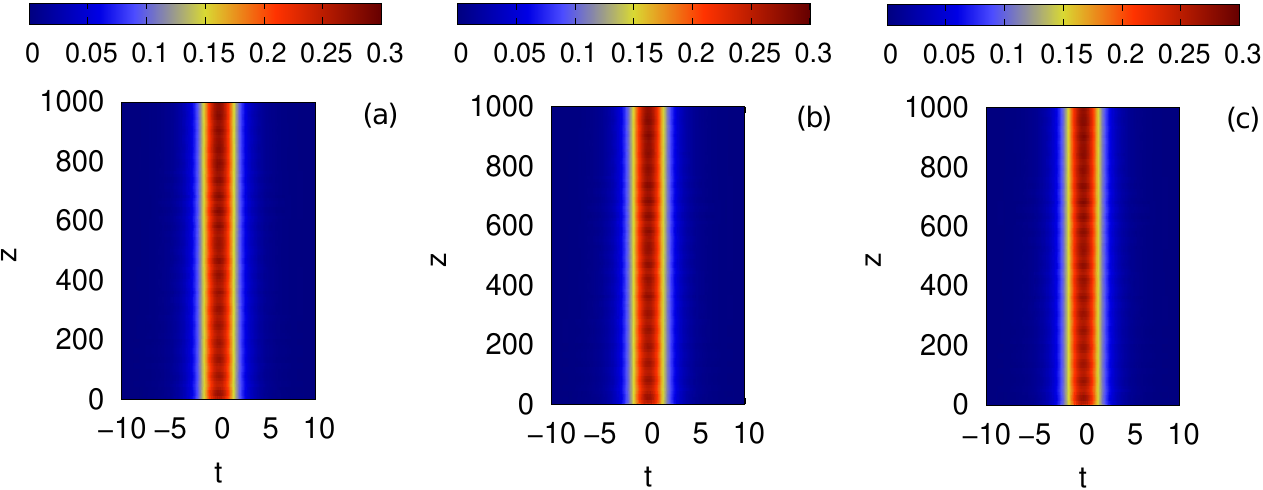} \caption{The perturbed evolution of components $|U_{1}|^{2}$, $|U_{2}|^{2}$
and $|U_{3}|^{2}$ of a stable symmetric soliton with $\alpha=1.6$,
$\lambda=0.5$ and $E=3$, is displayed in panels (a), (b) and (c),
respectively. The profiles were initially perturbed according to Eq.
(\ref{rd}).}
\label{F5} 
\end{figure*}

The single stationary FNLS equation, which corresponds to the decoupled
system(\ref{realU}) with $\lambda=0$, i.e., 
\begin{equation}
ku+\frac{1}{2}\left(-\frac{d^{2}}{dt^{2}}\right)^{\alpha/2}u-u^{3}=0.\label{single}
\end{equation}
gives rise to known scaling relations for its soliton solutions \citep{review,review2}.
Namely, varying $k$ leads to the following \emph{exact} relations
between $k$, the soliton's amplitude $A$ and its width $W$:
\begin{equation}
k\sim W^{-\alpha}\sim A^{2}.\label{scaling}
\end{equation}
Accordingly, the energy of the single-component soliton, $E=\int_{-\infty}^{+\infty}u^{2}(t)dt$,
scales as $E\sim A^{2}W$, i.e., $W\sim E/A^{2}$. The substitution
of this expression for $W$ in Eq. (\ref{scaling}) leads to the following
exact relation between $A^{2}$ and $E$:
\begin{equation}
A\sim E^{\alpha/2(\alpha-1)}.\label{A}
\end{equation}
In a detailed form, Eq. (\ref{A}) can be rewritten as
\begin{equation}
A=A_{0}(\alpha)E^{\alpha/2(\alpha-1)},\label{A2}
\end{equation}
where coefficient $A_{0}(\alpha)$ may be a function of LI $\alpha$,
but not of $E$. \\

In addition to that, it follows from Eqs. (\ref{A}) and (\ref{scaling})
that $k$ and $E$ are related by the following scaling,
\begin{equation}
k=k_{0}(\alpha)E^{\alpha/(\alpha-1)}.\label{k}
\end{equation}
The limit values of coefficients $A_{0}$ and $k_{0}$ in Eqs. (\ref{A2})
and (\ref{k}) are $A_{0}(\alpha=2)=1/2$ and $k_{0}(\alpha=2)=1/8$
(see Fig. \ref{F0}), which correspond to the classical soliton solutions
of the non-fractional NLS equation, with $\alpha=2$. On the other
hand, the singularity of relations (\ref{A2}) and (\ref{k}) in the
limit of $\alpha=1$ implies that the limit values of the coefficients
are $A_{0}(\alpha\rightarrow1)$, $k_{0}(\alpha\rightarrow1)\rightarrow0$.

\begin{figure*}[t]
\centering \includegraphics[width=0.8\textwidth]{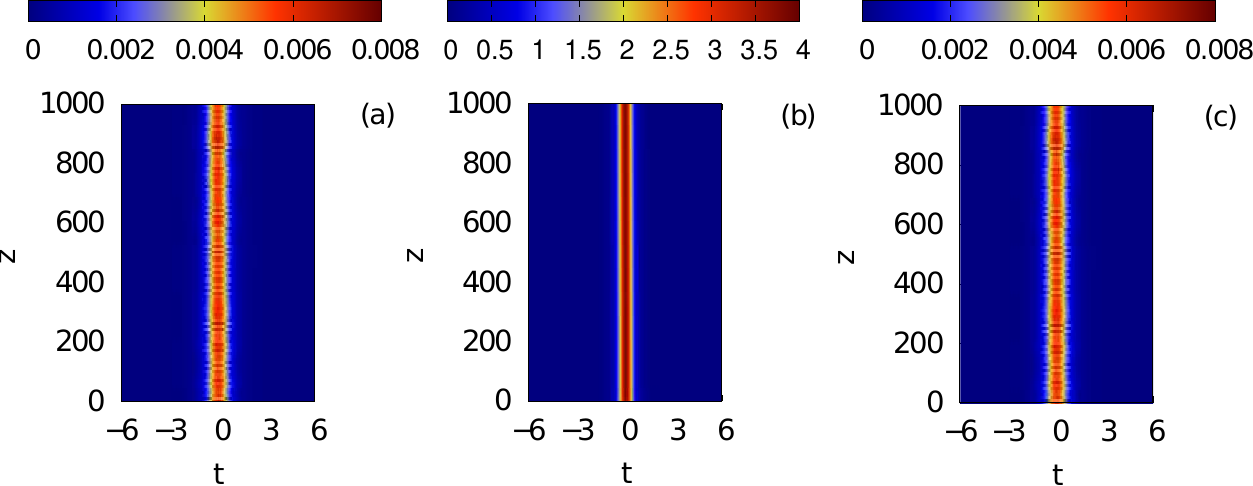} \caption{The perturbed evolution of components $|U_{1}|^{2}$, $|U_{2}|^{2}$
and $|U_{3}|^{2}$ of a stable asymmetric soliton for $\alpha=1.7$,
$\lambda=0.1$, $E=3$ and input $\Delta_{C}$, is displayed in panels
(a), (b) and (c) respectively.}
\label{F6} 
\end{figure*}

To illustrate these relations, in Fig. \ref{F0} we display three
different curves of $A_{0}$ and $k_{0}$ versus $\alpha$. The results
where obtained as numerically found ground-states of Eq. (\ref{single})
for $E=1.0$, $1.3$ and $1.5$, which identically coincide, in agreement
with Eqs. (\ref{A2}) and (\ref{k}). Thus, coefficients $A_{0}(\alpha)$
and $k_{0}(\alpha)$ are universal characteristics of the solitons
produced by the single FNLS equation.

\section{Spontaneous symmetry breaking (SSB) in three- \\
component ground-state (GS) solitons \label{Sec3}}

Numerical simulations of the system (\ref{EQ1}) were performed by
dint of the imaginary- and real-time propagation algorithms based
on the Fourier spectral method \citep{Yang_10}. The imaginary-time
simulations were used, as usual \citep{Bao}, to construct stationary
soliton solutions that represent the system's ground state (GS), which
minimizes the Hamiltonian (\ref{H}) for given total energy (\ref{Energy}).
Note that the symmetry of the underlying three-core system implies
that GSs represented by asymmetric solitons are degenerate, as mutually
equivalent GSs can be obtained from each other by cyclic transpositions
of subscripts $1,2,3$, cf. Refs. \citep{Salasnich_MP11,Mazzarella_PRA10,Santos_NL22,Miranda_PLA22}.
Then, the real-time simulations were employed to test stability of
the solitons and study their dynamical properties.

The imaginary-time simulations were initiated from the set of Gaussian
inputs $U_{j}(t,z=0)=B_{j}\exp(-t^{2})$, with $j=1$, $2$ and $3$.
The corresponding amplitudes $B_{j}$'s were obtained to produce three
initial conditions following small values of asymmetries (\ref{Theta}):
\begin{subequations}
\begin{equation}
\Delta_{A}\equiv\begin{cases}
\Theta_{1,2}=0.01,\\
\Theta_{2,3}=0.01,\\
\Theta_{1,3}=0.02,
\end{cases}\label{AS1}
\end{equation}
\begin{equation}
\Delta_{B}\equiv\begin{cases}
\Theta_{1,2}=0.01,\\
\Theta_{2,3}=-0.01,\\
\Theta_{1,3}=0,
\end{cases}\label{AS2}
\end{equation}
\begin{equation}
\Delta_{C}\equiv\begin{cases}
\Theta_{1,2}=-0.01,\\
\Theta_{2,3}=0.01,\\
\Theta_{1,3}=0.
\end{cases}\label{AS3}
\end{equation}
As shown below, these initial conditions can produce GS solitons with
strong asymmetry.

\begin{figure*}[htb]
\centering \includegraphics[width=0.8\textwidth]{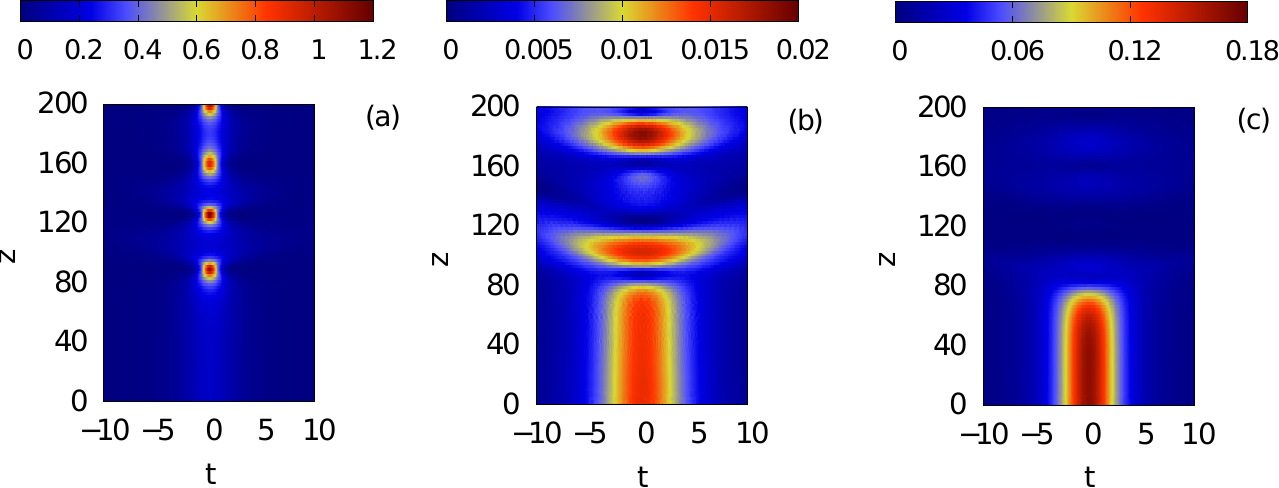} \caption{The perturbed evolution of components $|U_{1}|^{2}$, $|U_{2}|^{2}$
and $|U_{3}|^{2}$ of an unstable asymmetric soliton, with $\alpha=1.5$,
$\lambda=0.02$, $E=2$ and input $\Delta_{B}$, are displayed in
the panels (a), (b) and (c), respectively. }
\label{F7} 
\end{figure*}

In Fig. \ref{F1} we present generic examples of GS solitons. In particular,
panels \ref{F1}(a) and \ref{F1}(b) displays the profiles obtained
with two different values of $\lambda$, for LI $\alpha=1.8$ and
total power $E=3$. Naturally, the GS\ soliton in panel \ref{F1}(a),
supported by relatively strong linear coupling, with $\lambda=0.5$,
is symmetric, while the one in panel \ref{F1}(b), with much weaker
linear coupling ($\lambda=0.1$) is strongly asymmetric.

Further, to elucidate the effect of the initial asymmetry, Figs. \ref{F1}(b-d)
present asymmetric GS solitons obtained from different inputs (\ref{AS1})-(\ref{AS3}),
keeping the same values of $\alpha$, $\lambda$ and $E$. The configurations
displayed in these figures are drastically different, despite being
obtained with the same parameters. In Fig. \ref{F1}(b), input $\Delta_{A}$
produces identical (mutually symmetric) profiles $|u_{2}|^{2}$ and
$|u_{3}|^{2}$, while $|u_{1}|^{2}$ is different. Accordingly, in
this case the asymmetry is characterized by the combination 
\end{subequations}
 
\begin{equation}
E_{2}=E_{3}<E_{1},\label{E123}
\end{equation}
with the amplitudes of $u_{2}$ and $u_{3}$ which are much smaller
than that of $u_{1}$, see the insert in Fig. \ref{F1}(b).

The GS soliton resulting in Fig. \ref{F1}(d) from the input of type
$\Delta_{C}$ (Eq. (\ref{AS3})) is asymmetric too, but subject to
relation $E_{1}=E_{3}<E_{2}$, opposite to the one in Eq. (\ref{E123}).
The input of type $\Delta_{A}$ (see Eq. (\ref{AS1})) with the same
values of the system's parameters produces the same GS soliton as
in Fig. \ref{F1}(d), but with $u_{1}\longleftrightarrow u_{2}$.
The GS solitons the tri-core system with $\alpha=2$ (the usual non-fractional
GVD) are similar to those produced here with $\alpha=1.8$ (Figs.
\ref{F1}(a-d)), with the same symmetry and stability properties.

\begin{figure*}[htb]
\centering \includegraphics[width=0.9\textwidth]{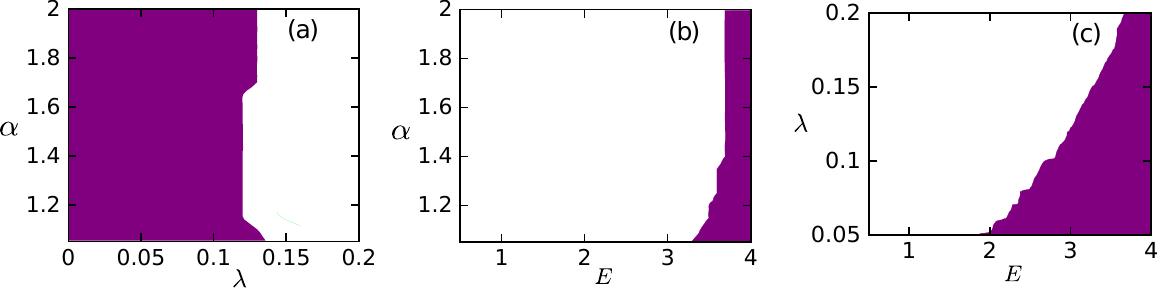}
\caption{Colored areas display bistability areas, in which stable asymmetric
solitons, produced by the inputs of types $\Delta_{A}$ and $\Delta_{C}$
(see Eqs. (\ref{AS1}) and (\ref{AS3})) coexist
in the parameter planes plotted here. Other parameters are fixed as
in Fig.  \ref{F4}.}
\label{FBs1} 
\end{figure*}

The GS solitons change abruptly under the action of FGVD, with smaller
values of LI, such as $\alpha=1.1$ (recall that all solitons are
destabilized by the collapse in the case of $\alpha\leq1$). For instance,
Fig. \ref{F1}(e) displays a fully symmetric GS soliton. In this case,
the symmetric profiles are wide, in comparison to the case of larger
LI, cf. Fig. \ref{F1}(a). As concerns asymmetric GS solitons, they
concentrate their energy in a single component, in agreement with
the asymmetry of the input. In particular, the inputs of types $\Delta_{A}$
and $\Delta_{C}$ produce, respectively, GS solitons in which virtually
all energy is concentrated in component $u_{1}$ or $u_{2}$, as can
be seen in Figs. \ref{F1}(f) and (h). Another characteristic feature
of the low-LI regime is a possibility of having a stable symmetric
profile in the case of weak linear coupling ($\lambda=0.1$), see
Fig. \ref{F1}(g).

In Fig. \ref{F3} we summarize the results for the GS soliton families
by means of plots for the asymmetry ratios (\ref{Theta}) versus the
linear-coupling constant $\lambda$, total energy $E$, and LI $\alpha$,
obtained from the same inputs labeled by symbols $\Delta_{A}$, $\Delta_{B}$
and $\Delta_{C}$ according to Eqs. (\ref{AS1})-(\ref{AS3}). Here
we focus the analysis on the configurations obtained with low LI values,
i.e., ones are that most different from those produced by the usual
(non-fractional) system, with $\alpha\rightarrow2$. Fig. \ref{F3}(a)
demonstrates abrupt change in $\Theta$ following the decrease of
$\lambda$, while $E$ and $\alpha$ are fixed. Thus we observe that
only asymmetric configurations are found in the regime of weak linear
coupling, for $\lambda\leq0.13$. In Fig. \ref{F3}(b) where the input
of the $\Delta_{B}$ type is used, the transition from asymmetric
configurations to the symmetric ones (the SSB bifurcation point) is
occurs at a still smaller value of the coupling constant, $\lambda=0.08$.
The curves $\Theta(\lambda)$ obtained with input $\Delta_{C}$ demonstrate,
in Fig. \ref{F3}(c), an abrupt transition, as in Fig. \ref{F3}(a).
However, in this configuration $u_{2}$ has higher energy than $u_{1}$,
producing $\Theta_{1,2}<0$. Note that, the bifurcation point on the
$\Theta(\lambda)$ curves are the same, indicating that the type of
the input does not effect the SSB bifurcation.

\begin{figure*}[t]
\centering \includegraphics[width=0.9\textwidth]{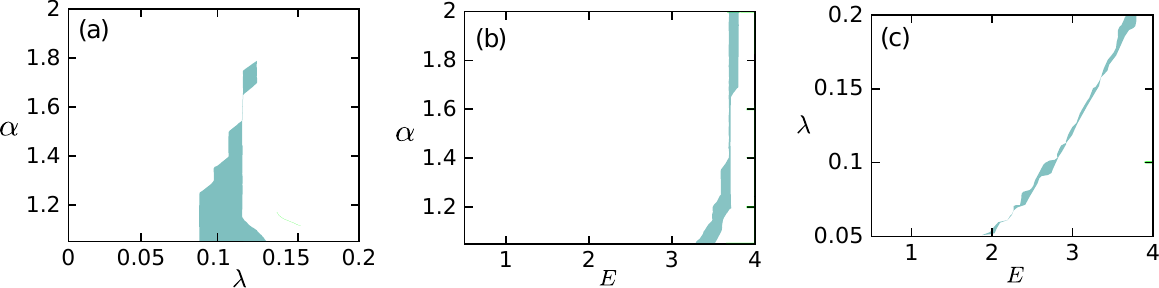}
\caption{Colored areas display tristability areas, in which two stable asymmetric
solitons and a stable symmetric one, produced by the inputs of types
$\Delta_{A}$, $\Delta_{C}$ and $\Delta_{B}$, respectively (see
Eqs. (\ref{AS1}) - (\ref{AS3})) coexist in the
parameter planes plotted here. Other parameters are fixed as in Fig.
\ref{F4}.}
\label{FBs2} 
\end{figure*}

To explore the effect of the total energy $E$, curves $\Theta(E)$
for the different inputs are plotted in Figs. \ref{F3}(d-f). It is
observed that, like in other SSB systems \citep{Peng}, the total
energy controls the SSB in the present system too. The onset of SSB
at the bifurcation point, exhibited by Figs. \ref{F3}(d-f), identifies
the bifurcation as one of the \textit{supercritical} type { (alias
the symmetry-breaking phase transition of the second kind \citep{bif}),
which implies that the symmetric states are destabilized by the bifurcation,
while the asymmetric ones emerge as stable states, no bistability
between symmetric and asymmetric solutions taking place.}

Completely novel results, that demonstrate the asymmetry ratios as
functions of LI, are presented in Figs. \ref{F3}(g-i). In particular,
a novel finding, reported in Figs. \ref{F3}(g) and (i), is that curves
$\Theta_{1,2}(\alpha)$ and $\Theta_{1,3}(\alpha)$ demonstrate two
distinct regions of asymmetry, separated by broad symmetry interval,
for the GS solitons produced by inputs $\Delta_{A}$ and $\Delta_{C}$.
On the other hand, input $\Delta_{B}$ produces solely fully symmetric
states, as shown in Fig. \ref{F3}(h).

To further summarize the findings, we have identified existence regions
for different states, in the relation to parameters $\lambda$, $E$
and $\alpha$. In Fig. \ref{F4}(a) we address the GS solitons obtained
with $E=3$ from input of type $\Delta_{A}$, varying LI $\alpha$
and the linear-coupling constant $\lambda$. In general, the plot
in the ($\lambda$, $\alpha$) plane shows that the asymmetric states
are naturally favored in the region of weak linear coupling . In Figs.
\ref{F4}(b) and (c), the same results are displayed, but now with
the use of input $\Delta_{B}$ or $\Delta_{C}$. In the former case,
the SSB area is somewhat larger.

In Figs. \ref{F4}(d-f), we address the effect of the total energy
on the GS solitons at different values of LI, while fixing the coupling
parameter as $\lambda=0.2$. The resulting diagrams show that, as
might be expected, the asymmetric states dominate at higher energies
(i.e., stronger self-focusing nonlinearity), and are weakly favored
by the decrease of LI. As in the previous cases, the diagram corresponding
to input $\Delta_{B}$ is slightly different (see Fig. \ref{F4}(b))
from those corresponding to $\Delta_{A}$ and $\Delta_{C}$.

Finally, in Fig. \ref{F4}(g-i) we investigated the combined effect
of the coupling strength, $\lambda$, and total energy, $E$, on the
GS solitons, fixing LI as $\alpha=1.7$. Naturally, the symmetric
GS is stabilized by the increase of $\lambda$ and destabilized by
the increase of $E$.

We tested the stability of GS solitons in direct real-time simulations,
using the perturbed input 
\begin{equation}
u_{1,2,3}(t,z=0)=[1+r_{1,2,3}(t)]u_{1,2,3}^{\mathrm{GS}}(t,z=0),\label{rd}
\end{equation}
where $u_{1,2,3}^{GS}(t,z=0)$ is the GS soliton obtained by the imaginary-time
propagation, and $r_{1,2,3}(t)$ is a random function produced by
the \textsl{rand} function of the \textsl{GNU-Octave} software, that
perturbs the initial condition. Here, the maximum of the random perturbation
corresponds to $\pm3\%$ of the unperturbed soliton's amplitude, i.e.,
the random function takes values $-0.03\leq r_{1,2,3}(t)\leq0.03$
with mean value $\langle r_{1,2,3}(t)\rangle\simeq0$. Using this
protocol, we tested the stability of the GS solitons obtained above.

Symmetric soliton exhibit dynamic stability in the tests. In Fig.
\ref{F5}, we illustrate this conclusion by displaying the evolution
of profiles $|U_{1,2,3}(t,z)|^{2}$ of a perturbed symmetric solitons
obtained for $\alpha=1.6$, $\lambda=0.5$ and $E=3$.

The tests of the evolution of the asymmetric solitons reveal stable
and unstable ones. Further, the simulations demonstrate that, generally,
the asymmetry solitons are more sensitive to disturbances. We conclude
that the asymmetric solitons produced by the inputs of types $\Delta_{A}$
and $\Delta_{C}$ (see Eqs. (\ref{AS1}) and (\ref{AS3})) demonstrate
stability. As an example, in Fig. \ref{F6} we present the perturbed
evolution of the asymmetric profiles obtained with $\alpha=1.7$,
$\lambda=0.1$, $E=3$ and $\Delta_{C}$. The profiles $|U_{1}|^{2}$
and $|U_{3}|^{2}$ exhibit variations, but the soliton as a whole
keeps its integrity, as shown by the fact that the evolution asymmetry
ratios $\Theta(z)$ keep constant values (not shown here in detail).

On the contrary to what is shown above, the perturbed evolution of
the asymmetric soliton generated by input $\Delta_{B}$ (see Eq. (\ref{AS2}))
demonstrates that the solitons of this type are completely unstable.
In this case, the initial disturbance in Eq. (\ref{rd}) is enough
to initiate apparent onset of instability, eventually leading to decay
of the soliton (delocalization of the wave fields). As an example,
we show in Fig. \ref{F7} the unstable perturbed evolution of an asymmetric
soliton obtained with $\alpha=1.5$, $\lambda=0.02$ and $E=2$. The
shapes of the three components start to change abruptly at $z\sim50$.
In particular, the profile $|U_{1}(t,z)|^{2}$, which had amplitude
$0.17$ at $z=0$, quickly increases its value to $1.1$ at $z=88$.
After this stage, the amplitude of $|U_{1}|^{2}$ decreases, performing
quasi-periodic oscillations. Simultaneously, component $U_{2}$ presents
slower oscillations, and $U_{3}$ suffers complete decay, showing
the remaining energy $E_{3}=0.014$ at $z=120$.

Figs. \ref{F1}(b) and \ref{F1}(d) demonstrate the coexistence of
two different asymmetric solutions obtained with the same system's
parameters $\alpha$ and $\lambda$, and equal energies $E$. These
profiles, generated by different inputs, of types $\Delta_{A}$ and
$\Delta_{C}$ (see Eqs. (\ref{AS1}) and (\ref{AS3})), are stable,
indicating the phenomenon of bistability. The coexistence regions
of different stable asymmetric profiles are shown in Fig. \ref{FBs1}.
We also investigated the tristability, coexistence of three different
types of stable solitons (tristability), \textit{viz}., two asymmetric
ones and a symmetric soliton, also found for the same parameters.
An example of this is provided a trio of solitons shown in Figs. \ref{F1}(f-h).
Fig. \ref{FBs2} displays parameter areas where the tristability is
found in the planes of ($\lambda$, $\alpha$), ($E$, $\alpha$)
and ($E$, $\lambda$).

\begin{figure}[tb]
\centering \includegraphics[width=0.7\columnwidth]{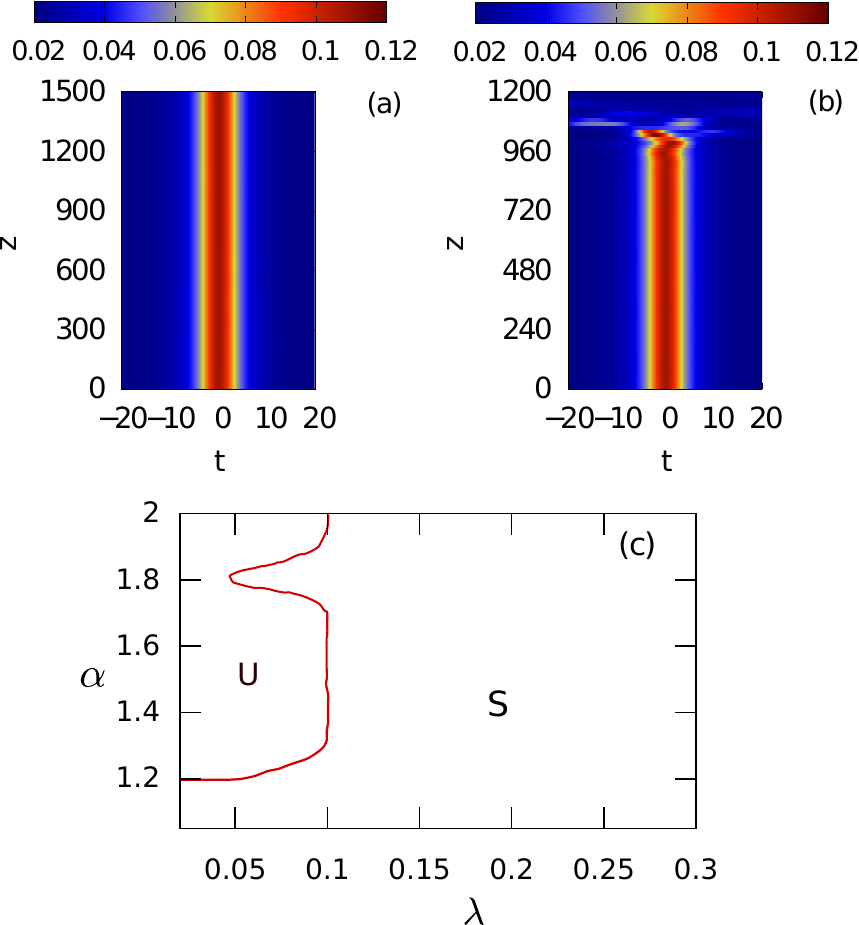} \caption{The evolution of the component $|U_{3}|^{2}$ of an (un)stable vortex
soliton (\ref{uj}) for $\alpha=1.2$ and $\lambda=0.1$
(a) or $\lambda=0.05$ (b), as produced by simulations of Eq. (\ref{EQ1}).
The total energy of the vortex soliton is $E=3$. The evolution of
other components, $U_{1}$ and $U_{2}$, is similar. (c) The stability
diagram for the vortex solitons with $E=3$. They are stable and unstable
in areas S and U, respectively.}
\label{F8} 
\end{figure}

\begin{table*}[ht]
\centering %
\begin{tabular}{c|ccc}
\hline 
 & Initial condition of asymmetry  & $\lambda_{c}$ ($\alpha=1.6$ and $E=3$)  & $E_{c}$ ($\alpha=1.6$ and $\lambda=0.13$) \tabularnewline
\hline 
\multirow{3}{*}{\makecell[lc]{\rotatebox[origin=c]{90}{Weak}}} & $\Delta_{A}\,(\Theta_{12}=0.01,\,\Theta_{23}=0.01,\,\Theta_{13}=0.02)$  & $0.13$  & $2.9$ \tabularnewline
 & $\Delta_{B}\,(\Theta_{12}=0.01,\,\Theta_{23}=-0.01,\,\Theta_{13}=0)$  & $0.12$  & $2.9$ \tabularnewline
 & $\Delta_{C}\,(\Theta_{12}=-0.01,\,\Theta_{23}=0.01,\,\Theta_{13}=0)$  & $0.13$  & $2.9$ \tabularnewline
\hline 
\multirow{3}{*}{\makecell[lc]{\rotatebox[origin=c]{90}{Strong}}} & $\Delta_{A'}\,(\Theta_{12}=0.2,\,\Theta_{23}=0.4,\,\Theta_{13}=0.6)$  & $0.24$  & $2.3$ \tabularnewline
 & $\Delta_{B'}\,(\Theta_{12}=1,\,\Theta_{23}=-1,\,\Theta_{13}=0)$  & $0.12$  & $3.1$ \tabularnewline
 & $\Delta_{C'}\,(\Theta_{12}=-0.99,\,\Theta_{23}=0.99,\,\Theta_{13}=0)$  & $0.36$  & $1.9$ \tabularnewline
\cline{2-4} \cline{3-4} \cline{4-4} 
\end{tabular}\caption{Comparison between the critical values of interaction $\lambda_{c}$
and energy $E_{c}$, obtained with weak ($\Delta_{A}$, $\Delta_{B}$
and $\Delta_{C}$) and strong ($\Delta_{A}'$, $\Delta_{B}'$ and
$\Delta_{C}'$) initial asymmetry conditions. The critical values
define the onset of the SSB region.}
\label{tab1} 
\end{table*}

{The above analysis is focused on slightly asymmetric initial conditions
(\ref{AS1})-(\ref{AS3}), aiming to minimize the effect of the input's
asymmetry and highlight the role of the inner evolution governed by
system (\ref{EQ1}). Extensive simulations, conducted for various
degrees of asymmetry of the initial conditions, produce similar results
for GS solitons. Nevertheless, SSB regions may differ, depending on
the degree of the initial asymmetry. To investigate this feature,
we analyzed the shapes of the GS solitons obtained for three strong
asymmetric inputs, \textit{viz.}, $\Delta_{A'}\equiv\Theta_{12}=0.2,\,\Theta_{23}=0.4,\,\Theta_{13}=0.6$;
$\Delta_{B'}\equiv\Theta_{12}=1,\,\Theta_{23}=-1,\,\Theta_{13}=0$;
and $\Delta_{C'}\equiv\Theta_{12}=-0.99,\,\Theta_{23}=0.99,\,\Theta_{13}=0$,
cf. Eqs. (\ref{AS1})-(\ref{AS3}). The respective findings are compared
to those obtained from the weakly asymmetry inputs ($\Delta_{A}$,
$\Delta_{B}$ and $\Delta_{C}$), as given by Eqs. (\ref{AS1})-(\ref{AS3}).
Table \ref{tab1} summarizes the respective results, presenting the
corresponding critical values of the coupling constant and energy,
$\lambda_{c}$ and $E_{c}$, respectively. As above, SSB takes place
in the GS solitons at $\lambda<\lambda_{c}$ or $E>E_{c}$. We conclude
that, while SSB occurs in similar regions for the GS solitons produced
with the weak input's asymmetry, the situation is different in the
case of the strong initial asymmetry. For instance, only asymmetric
solitons are found at $\lambda<0.24$ and $E>2.3$ when $\Delta_{A'}$
is considered. Differently, in the same settings but using $\Delta_{C'}$,
the asymmetric states are present at $\lambda<0.36$ and $E>1.9$.
Therefore, we conclude that, quite naturally, the strong initial asymmetry
significantly catalyzes the onset of SSB.}

\section{Vortex solitons\label{ivort}}

\subsection{Stability}

In two-dimensional models with the fractional diffraction, vortices
and their stability were studied in many works \citep{vort2}-\citep{Yingji2},
\citep{Hidetsugu}. In the present setting, following Ref. \citep{Sigler},
the vortex soliton is defined as a set of three solitons in the coupled
cores, with phase shifts $2\pi/3$ between them: 
\begin{equation}
U_{j}(t,z)=u_{j}(t)\exp\left[i\frac{2\pi\sigma}{3}(j-1)+ikz\right],~j=1,2,3.\label{uj}
\end{equation}
The vorticity, with sign (winding number) $\sigma=\pm1$, is represented
by the fact that the total phase gain produced by the round trip comprising
the three cores is $\Delta\phi=2\pi\sigma$, according to ansatz (\ref{uj}).

If functions $u_{j}(t)$ are real, the substitution of ansatz (\ref{uj})
in Eqs.~\ref{realU} admits solutions with 
\begin{equation}
u_{1}=u_{2}=u_{3}\equiv u(t),\label{uuu}
\end{equation}
satisfying the single real equation:
\begin{equation}
(k-\lambda)u+\frac{1}{2}\left(-\frac{d^{2}}{dt^{2}}\right)^{\alpha/2}u-u^{3}=0.\label{U}
\end{equation}
Equation (\ref{U}) is the standard fractional one, which was solved
in many works. The solution is characterized by the total energy (\ref{Energy}),
which reduces to
\begin{equation}
E=\int_{-\infty}^{+\infty}\sum_{j=1,2,3}\left\vert u_{j}(t)\right\vert ^{2}dt\equiv3\int_{-\infty}^{+\infty}u^{2}(t)dt.\label{Evort}
\end{equation}
Our main objective here is to identify a stability area of vortex
solitons (\ref{uj}) in the parameter planes of $\left(\lambda,\alpha\right)$
or $\left(E,\alpha\right)$.

To investigate the stability of the vortex modes defined as per Eq.
(\ref{uj}) with $\sigma=+1$ (the case of $\sigma=-1$ introduces
no difference), the input profile was chosen as the symmetric GS of
the decoupled system, i.e., $u_{1}(t)=u_{2}(t)=u_{3}(t)=u^{\mathrm{GS}}(t,\lambda=0)$.

Numerical simulations show the existence of stable and unstable vortex
soliton. Fig. \ref{F8}(a) presents an example of stable evolution
of a vortex soliton with total energy $E=3$, in the weakly coupled
system with $\lambda=0.1$ and LI $\alpha=1.2$. The simulations demonstrate
stable propagation of this vortex state. It loses stability, following
the decrease of the linear-coupling constant $\lambda$. In Fig. \ref{F8}(b),
an example of the unstable evolution is displayed, demonstrating the
destruction of the vortex soliton (decay into radiation) close to
$z=1050$.

The stability area for the vortex solitons is displayed in Fig. \ref{F8}(c).
It is seen that the stability of the vortex states is favored by low
values of LI $\alpha$. In particular, {the vortex modes are stable
in the interval of $1<\alpha<1.2$}.

\begin{figure}[tb]
\centering \includegraphics[width=0.7\columnwidth]{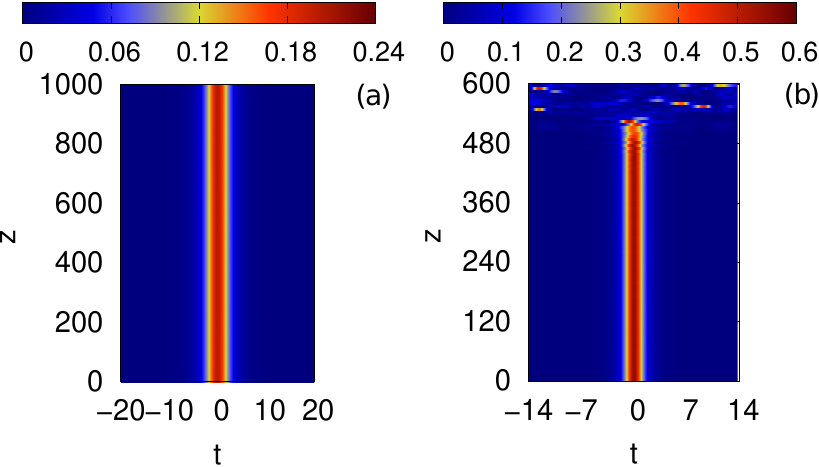} \caption{The evolution of component $|U_{3}|^{2}$ of the vortex solitons for$\lambda=0.3$
and $\alpha=1.5$. The total energy of the stable soliton in (a) is
$E=3$, and of the unstable one in (b) is $E=4.2$.}
\label{F9} 
\end{figure}

We also addressed the effect of the total energy of the vortex solitons
on their stability. As an example, Fig. \ref{F9} demonstrates the
evolution of the vortex solitons in the regime of relatively strong
coupling ($\lambda=0.3$), with LI $\alpha=1.5$, and energies $E=3$
in (a) and $E=4.2$ in (b). The simulations demonstrate the stability
of the former vortex soliton and instability of the latter one. Further
simulations demonstrate that the critical energy, above which the
vortex soliton develops the instability, is nearly independent of
LI. For example, setting $\lambda=0.1$, we conclude that the critical
energy is $E\approx3$ in the entire interval of the LI values, $1<\alpha\leq2$.
Similarly, setting $\lambda=0.3$, we conclude that the critical energy,
$E\approx3.3$, again keeps this nearly a constant value in the same
interval.

{ Unlike the GS solitons, the SSB effect was not found in the vortex
ones at all values of the parameters for which the analysis was performed
in this work. Actually, this fact stresses the exceptional robustness
of the vortex solitons, which are protected by their topological charge
against the symmetry breaking.}

\subsection{Mobility of vortex solitons}

\begin{figure}[bt]
\centering \includegraphics[width=0.8\columnwidth]{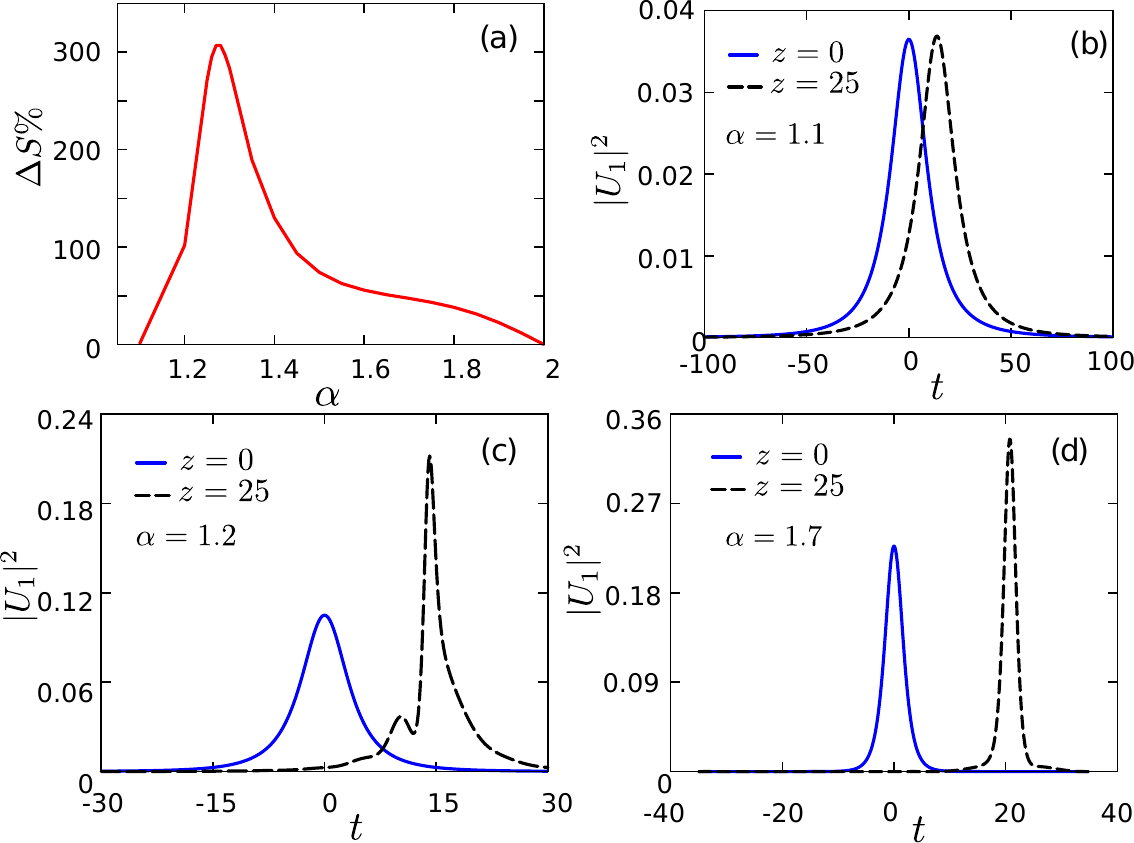} \caption{(a) Relative difference between static and moving vortex (at $z=25$)
mode peaks versus $\alpha$. Comparison between the vortex modes $|U_{1}(t,z=0)|^{2}$
and $|U_{1}^{b}(t,z=25)|^{2}$ for $\alpha=1.1$ (b), $\ \alpha=1.2$
(c) and $\alpha=1.7$ (d). }
\label{F11} 
\end{figure}

The FGVD destroys the usual Galilean invariance of the NLS equation,
therefore mobility of fractional solitons is a nontrivial issue \citep{review,review2}.
In this work, we address the mobility of vortex solitons, as their
internal structure may make them more sensitive to the motion than
in the case of the GS solitons considered above.

By means of direct simulations, we studies results of the application
of a boost to the stable vortex soliton, 
\begin{equation}
U_{j}^{b}(t,z)\rightarrow U_{j}^{b}(t,z)=U_{j}(t,z)\exp(it),\label{boost}
\end{equation}
with a fixed boost parameter corresponding to $\chi=-1$ in the general
expression for the boost factor, $\exp\left(-i\chi t\right)$.

The mobility effect is quantified by the relative difference of the
peak power between the moving and quiescent (static) vortex solitons,
\begin{equation}
\Delta S=\dfrac{\left\vert \text{max}|U_{1}^{b}(t,z)|^{2}-\text{max}|U_{1}(t,z=0)|^{2}\right\vert }{\text{max}|U_{1}(t,z=0)|^{2}}.\label{S}
\end{equation}
Figure \ref{F11}(a) shows the relative difference (\ref{S}) as a
function of LI $\alpha$, obtained for a fixed linear-coupling constant,
$\lambda=0.05$. The plot identifies the (relatively \ low) value
of LI, $\alpha=1.27$, at which the FGVD produce the largest nontrivial
effect of the mobility. Although $\Delta S$ is small at $\alpha<1.12$,
it is different from zero: for instance, $\Delta S\left(\alpha=1.10\right)=1.14\%$.

The temporal structures of the quiescent and moving vortex solitons
are compared in Figs. \ref{F11}(a-c). At $\alpha=1.1$ the boosted
vortex soliton moves slowly, featuring a small deformation in the
profile. At larger values of LI, such as $\alpha=1.2$, a large deformation
of the vortex is observed, showing partial fragmentation of the temporal
profile, initially with a weak peak emerging to the left of the main
one. In all cases, changes in moving profiles are independent of the
linear-coupling constant $\lambda$. This is explained by the fact
that the temporal profiles of all the components of the vortex solitons
remain identical, as in Eq. (\ref{uuu}).

\section{Conclusion \label{Sec4}}

In the framework of the tri-core optical system combining the intra-core
self-focusing and FGVD (fractional group-velocity dispersion), which
is modeled by the system of linearly coupled FNLS (fractional nonlinear
Schrödinger) equations, we have performed the analysis of GS (ground-state)
three-component solitons, and vortex solitons with the triangular
(prismatic) structure. For the GS solitons, the SSB (spontaneous-symmetry-breaking)
phenomenology is investigated in detail, as the functions of the LI
(Lévy index), linear-coupling constant, and total energy of the three-component
solitons. The existence diagrams for symmetric and asymmetric GS solitons
are plotted, and bifurcation diagrams for SSB are constructed, with
the conclusion that the bifurcation is of the supercritical type.
As usual, the SSB is driven by the increase of the nonlinearity and
decrease of the linear coupling between the cores. The existence and
stability diagrams are produced for vortex solitons too. Unlike the
GS solitons, the vortices do not feature SSB. The mobility of the
vortex solitons in the fractional medium, imposed by the application
of the boost, is studied too.

As an expansion of the work, it may be interesting to study solitons
and SSB phenomenology in them, in the framework of a planar, rather
than triangular, three-core fractional system. A challenging objective
is to develop the FGVD model for nonlinear multi-core systems with
the fractional dispersion.

\section*{Acknowledgments}

The authors acknowledge the financial support of the Brazilian agencies
CNPq (\#306105/2022-5) and FAPEG. This work was also performed as
part of the Brazilian National Institute of Science and Technology
(INCT) for Quantum Information (\#465469/2014-0). The work of BAM
was supported, in a part, by the Israel Science Foundation through
grant No. 1287/17.


\begin{thebibliography}{99}
\bibitem{ZMNP} V.E. Zakharov, S.V. Manakov, S.P. Novikov, L.P. Pitaevskii,
\textit{Theory of Solitons: The Inverse Problem Method }(Nauka Publishers,
Moscow, 1980) (English translation: Consultants Bureau, New York,
1984).

\bibitem{Dauxois} T. Dauxois, M. Peyrard, \textit{Physics of Solitons}
(Cambridge University Press, Cambridge, 2006)

\bibitem{Agrawal_13} G.P. Agrawal, \textit{Nonlinear Fiber Optics}
(Elsevier, 2013).

\bibitem{Kivshar_03} Y.S. Kivshar, G.P. Agrawal, \textit{Optical
Solitons} (Elsevier, 2003).

\bibitem{Zakharov_JETP72} V.E, Zakharov, \textit{Collapse of Langmuir
Waves}, J. Exp. Theor. Phys. \textbf{35} (1972) 908.

\bibitem{Ichikawa} Y.H. Ichikawa, \textit{Topics on solitons in plasmas},
Physica Scripta. \textbf{20} (1979) 296-305

\bibitem{Pethick_08} C.J. Pethick, H. Smith, \emph{Bose--Einstein
Condensation in Dilute Gase}s (Cambridge University Press, 2008).

\bibitem{Pitaevskii_03} L.P. Pitaevskii, S. Stringari, \textit{Bose-Einstein
Condensation} (Clarendon Press, 2003).

\bibitem{Laksh} M. Lakshmanan, \textit{The fascinating world of the
Landau--Lifshitz--Gilbert equation: an overview}, Phil. Trans. R.
Soc. A \textbf{369} (2011) 1280--1300

\bibitem{surface-waves} N.E. Huang, Z. Shen, S.R. Long, \textit{A
new view of nonlinear water waves: The Hilbert spectrum}, Ann. Rev.
Fluid Mech. \textbf{31} (1999) 417-457

\bibitem{surface-waves2} M.F. Gobbi, J.T. Kirby, G. Wei, \textit{A
fully nonlinear Boussinesq model for surface waves}, J. Fluid Mech.
\textbf{405} (2000) 181-210

\bibitem{Maugin} G.A. Maugin, \textit{Nonlinear surface waves and
solitons}, Eur. Phys. J. Special Topics, \textbf{147} (2007) 209--230

\bibitem{SY} J. Satsuma, N. Yajima, \textit{Initial value problems
of one-dimensional self-modulation of nonlinear waves in dispersive
media}, Suppl. Prog. Theor. Phys. \textbf{55} (1974) 284-306

\bibitem{Jensen} S.M. Jensen, \textit{The nonlinear coherent coupler},
IEEE J. Quantum Electron \textbf{18} (1982) 1580--1583

\bibitem{Maier} A.A. Maier, \textit{Optical transistors and bistable
devices utilizing nonlinear transmission of light in systems with
unidirectional coupled waves}, Sov. J. Quantum Electron \textbf{12}
(1982) 1490--1494

\bibitem{Trillo} M. Romagnoli, S. Trillo, and S. Wabnitz, \textit{Soliton
switching in nonlinear couplers}, Opt. Quantum Electron \textbf{24}
(1992) S1237--S1267

\bibitem{Peng} B.A. Malomed, \textit{A variety of dynamical settings
in dual-core nonlinear fibers}, In: \textit{Handbook of Optical Fibers},
Vol. 1, (G.-D. Peng, Editor: Springer, Singapore, 2019) pp. 421-474

\bibitem{Akhmediev_JOS94} N.N. Akhmediev, A.V. Buryak, \textit{Soliton
states and bifurcation phenomena in three-core nonlinear fiber couplers},
J. Opt. Soc. Am. B \textbf{11} (1994) 804

\bibitem{Arik} A. Gubeskys, B.A. Malomed, \textit{Solitons in a system
of three linearly coupled fiber gratings}, Eur. Phys. J. D \textbf{28}
(2004) 283-299

\bibitem{Agrawal} S. Mumtaz, R.J. Essiambre, G.P. Agrawal, \textit{Nonlinear
propagation in multimode and multicore fibers: generalization of the
Manakov equations}, J. Lightwave Technol. \textbf{31} (2013) 398--406

\bibitem{ultrahigh} R.G.H. van Uden, R. Amezcua Correa, E. Antonio
Lopez, F.M. Huijskens, C. Xia, G. Li, A. Schülzgen, H. de Waardt,
A.M.J. Koonen, C.M. Okonkwo, \textit{Ultra-high-density spatial division
multiplexing with a few-mode multicore fibre}, Nature Phot. \textbf{8}
(2014) 865--870

\bibitem{Wabnitz} I.S. Chekhovskoy, O.V. Shtyrina, S. Wabnitz, M.P.
Fedoruk, \textit{Finding spatiotemporal light bullets in multicore
and multimode fibers}, Opt. Express \textbf{28} (2020) 7817--7828

\bibitem{Wise} L.G. Wright, F.O. Wu, D.N. Christodoulides, F.W. Wise,
\textit{Physics of highly multimode nonlinear optical systems}, Nature
Phys. \textbf{18} (2022) 1018-1030

\bibitem{Ignac} V.H. Nguyen, L.X.T. Tai, I. Bugar, M. Longobucco,
R. Buzcynski, B.A. Malomed, M. Trippenbach, \textit{Reversible ultrafast
soliton switching in dual-core highly nonlinear optical fibers}, Opt.
Lett. \textbf{45} (2020) 5221-5224

\bibitem{Dong} T. Mayteevarunyoo, B.A. Malomed, G. Dong, \textit{Spontaneous
symmetry breaking in a nonlinear double-well structure}, Phys. Rev.
A \textbf{78} (2008) 53601

\bibitem{Shatrughna} S. Kumar, P. Li, L. Zeng, J. He, B.A. Malomed,
\textit{A solvable model for symmetry-breaking phase transitions},
Sci. Rep. \textbf{13}(2023) 13768

\bibitem{Acus} A. Acus, B.A. Malomed, Y. Shnir, \textit{Spontaneous
symmetry breaking of binary fields in a nonlinear double-well structure},
Physica D\textbf{\ 241} (2012) 987

\bibitem{Laskin_PLA00} N. Laskin, \textit{Fractional quantum mechanics
and Lévy path integrals}, Phys. Lett. A 268 (2000) 298

\bibitem{Laskin_PRA02} N. Laskin, \textit{Fractional Schrödinger
equation}, Phys. Rev. E \textbf{66} (2002) 56108

\bibitem{GuoXu} X. Guo, M. Xu, \textit{Some physical applications
of fractional Schrödinger equation}, J. Math. Phys. \textbf{47} (2006)
082104

\bibitem{Laskin_18} N. Laskin, \textit{Fractional Quantum Mechanics}
(World Scientific, Singapore, 2018).

\bibitem{Longhi_OL15} S. Longhi, \textit{Fractional Schrödinger equation
in optics}, Opt. Lett. \textbf{40} (2015) 1117

\bibitem{Shilong} S. Liu, Y. Zhang, B.A. Malomed, E. Karimi, \textit{Experimental
realisations of the fractional Schrödinger equation in the temporal
domain}, Nature Comm. \textbf{14} (2023) 222

\bibitem{Zhang_CNSNS17} L. Zhang, Z. He, C. Conti, Z. Wang, Y. Hu,
D. Lei, Y. Li, and D. Fan, \textit{Modulational instability in fractional
nonlinear Schrödinger equation}, Commun. Nonlinear Sci. Numer. Simul.
\textbf{48} (2017) 531

\bibitem{Zhong_PRE16} W.-P. Zhong, M.R. Beli\'{c}, B.A. Malomed,
Y. Zhang, T. Huang, \textit{Spatiotemporal accessible solitons in
fractional dimensions}, Phys. Rev. E \textbf{94} (2016) 012216

\bibitem{Zhong_AP16} W.-P. Zhong, M. Beli\'{c}, Y. Zhang, \textit{Accessible
solitons of fractional dimension}, Ann. Phys. (N. Y). \textbf{368}
(2016) 110

\bibitem{Secchi_AA14} S. Secchi, M. Squassina, \textit{Soliton dynamics
for fractional Schrödinger equations}, Appl. Anal. \textbf{93} (2014)
1702

\bibitem{Hong_N17} Y. Hong, Y. Sire, \textit{A new class of traveling
solitons for cubic fractional nonlinear Schrödinger equations}, Nonlinearity
\textbf{30} (2017) 1262

\bibitem{Chen_PRE18} M. Chen, S. Zeng, D. Lu, W. Hu, Q. Guo, \textit{Optical
solitons, self-focusing, and wave collapse in a space-fractional Schrödinger
equation with a Kerr-type nonlinearity}, Phys. Rev. E \textbf{98}
(2018) 022211

\bibitem{Wang_EPL18} Q. Wang, J. Li, L. Zhang, W. Xie, \textit{Hermite-gaussian--like
soliton in the nonlocal nonlinear fractional Schrödinger equation},
EPL \textbf{122} (2018) 64001

\bibitem{HS} H. Sakaguchi, B.A. Malomed, \textit{Two-dimensional
solitons in second-harmonic-generating media with fractional diffraction},
Physica D \textbf{467} (2024) 134242

\bibitem{Huang_OL16} C. Huang, L. Dong, \textit{Gap solitons in the
nonlinear fractional Schrödinger equation with an optical lattice},
Opt. Lett. \textbf{41} (2016) 5636

\bibitem{Xiao_OE18} J. Xiao, Z. Tian, C. Huang, L. Dong, \textit{Surface
gap solitons in a nonlinear fractional Schrödinger equation}, Opt.
Express \textbf{26} (2018) 2650

\bibitem{Li_OE20} P. Li, B.A. Malomed, D. Mihalache, \textit{Metastable
soliton necklaces supported by fractional diffraction and competing
nonlinearities}, Opt. Express \textbf{28} (2020) 34472

\bibitem{Qiu_CSF20} Y. Qiu, B.A. Malomed, D. Mihalache, X. Zhu, X.
Peng, Y. He, \textit{Stabilization of single-and multi-peak solitons
in the fractional nonlinear Schrödinger equation with a trapping potential},
Chaos, Solitons \& Fractals \textbf{140} (2020) 110222

\bibitem{Zeng_CSF21} L. Zeng, D. Mihalache, B.A. Malomed, X. Lu,
Y. Cai, Q. Zhu, J. Li, \textit{Families of fundamental and multipole
solitons in a cubic-quintic nonlinear lattice in fractional dimension},
Chaos, Solitons \& Fractals \textbf{144} (2021) 110589

\bibitem{Zeng_OL19} L. Zeng, J. Zeng, \textit{Families of fundamental
and multipole solitons in a cubic-quintic nonlinear lattice in fractional
dimension}, Opt. Lett. \textbf{44} (2019) 2661

\bibitem{Santos_CJP24} M.C.P. dos Santos, B.A. Malomed, W.B. Cardoso,
\textit{Solitons supported by a self-defocusing trap in a fractional-diffraction
waveguide}, Chin. J. Phys. \textbf{89} (2024) 1474-1482

\bibitem{Zeng_CP20} L. Zeng, J. Zeng, \textit{Preventing critical
collapse of higher-order solitons by tailoring unconventional optical
diffraction and nonlinearities}, Commun. Phys. \textbf{3} (2020) 26

\bibitem{Wang_JO20} Q. Wang, G. Liang, \textit{Vortex and cluster
solitons in nonlocal nonlinear fractional Schrödinger equation}, J.
Optics \textbf{22}(2020) 055501

\bibitem{Li_CSF20} P. Li, B.A. Malomed, D. Mihalache, \textit{Vortex
solitons in fractional nonlinear Schrödinger equation with the cubic-quintic
nonlinearity}, Chaos, Solitons \& Fractals \textbf{137} (2020) 109783

\bibitem{review} B.A. Malomed, \textit{Optical solitons and vortices
in fractional media: a mini-review of recent results}, Photonics \textbf{8}
(2021) 353

\bibitem{review2} B.A. Malomed, \textit{Basic fractional nonlinear-wave
models and solitons, Optical solitons and vortices in fractional media:
A mini-review of recent results}, Chaos \textbf{34} (2024) 022102

\bibitem{Zeng_CSF20} L. Zeng, J. Zeng, \textit{Fractional quantum
couplers}, Chaos, Solitons \& Fractals \textbf{140} (2020) 110271

\bibitem{Zeng_ND21} L. Zeng, J. Shi, X. Lu, Y. Cai, Q. Zhu, H. Chen,
H. Long, J. Li, \textit{Stable and oscillating solitons of $\mathcal{PT}$-symmetric
couplers with gain and loss in fractional dimension}, Nonlinear Dyn.
\textbf{103} (2021) 1831

\bibitem{Strunin} D.V. Strunin, B. A. Malomed, \textit{Symmetry-breaking
transitions in quiescent and moving solitons in fractional couplers},
Phys. Rev. E \textbf{107} (2023) 064203

\bibitem{Sigler} A. Sigler, B.A. Malomed, D.V. Skryabin, \textit{Localized
states in a triangular set of linearly coupled complex Ginzburg-Landau
equations}, Phys. Rev. E \textbf{74} (2006) 066604



\bibitem{Riesz} M. Cai, C.P. Li, \textit{On Riesz derivative}, Fract.
Calc. Appl. Anal. \textbf{22} (2019) 287--301

\bibitem{Mandelbrot} B.B. Mandelbrot, \textit{The Fractal Geometry
of Nature} (W. H. Freeman, New York, 1982).

\bibitem{Hidetsugu} H. Sakaguchi, B. A. Malomed, \textit{One- and
two-dimensional solitons in spin-orbit-coupled Bose-Einstein condensates
with fractional kinetic energy}, J. Phys. B: At. Mol. Opt. Phys. \textbf{55}
(2022) 155301

\bibitem{Chen_CHAOS20} J. Chen, J. Zeng, \textit{Spontaneous symmetry
breaking in purely nonlinear fractional systems}, Chaos \textbf{30}
(2020) 063131

\bibitem{Salasnich_MP11} L. Salasnich, B.A. Malomed, \textit{Spontaneous
symmetry breaking in linearly coupled disk-shaped Bose-Einstein condensates},
Mol. Phys. \textbf{109} (2011) 2737-2745

\bibitem{DUO_CMA16} S. Duo, Y. Zhang, \textit{Mass-conservative Fourier
spectral methods for solving the fractional nonlinear Schr\'{'}{o}dinger
equation}, Comput. Math. with Appl. \textbf{71} (2016) 2257

\bibitem{Jeng_JMP10} M. Jeng, S.-L.-Y. Xu, E. Hawkins, J.M. Schwarz,
\textit{On the nonlocality of the fractional Schrödinger equation},
J. Math. Phys. \textbf{51} (2010) 62102

\bibitem{Luchko_JMP13} Y. Luchko, \textit{Fractional Schrödinger
equation for a particle moving in a potential well}, J. Math. Phys.
\textbf{54} (2013) 12111

\bibitem{Yang_10} J. Yang, \textit{Nonlinear Waves in Integrable
and Nonintegrable Systems} (SIAM, Philadelphia, 2010).

\bibitem{Bao} W.Z. Bao, Q. Du, \textit{Computing the ground state
solution of Bose-Einstein condensates by a normalized gradient flow},
SIAM J. Sci. Comp. \textbf{25} (2004) 1674-1697

\bibitem{Santos_NL22} M.C.P. dos Santos, W. B. Cardoso, \textit{Localization
of light waves in self-defocusing fractional systems confined by a
random potential}, Nonlinear Dyn. \textbf{112} (2024) 2209-2217

\bibitem{Mazzarella_PRA10} G. Mazzarella, L. Salasnich, \textit{Spontaneous
symmetry breaking and collapse in bosonic Josephson junctions}, Phys.
Rev. A \textbf{82} (2010) 33611

\bibitem{Miranda_PLA22} B.M. Miranda, M.C.P. dos Santos, W.B. Cardoso,
\textit{Symmetry breaking in Bose-Einstein condensates confined by
a funnel potential}, Phys. Lett. A \textbf{452} (2022) 128453

\bibitem{bif} G. Iooss, D.D. Joseph, \textit{Elementary Stability
Bifurcation Theory} (Springer-Verlag, New York, 1980).

\bibitem{vort2} X. Yao, X. Liu, \textit{Off-site and on-site vortex
solitons in space-fractional photonic lattices}, Opt. Lett. \textbf{43}
(2018) 5749--5752

\bibitem{vort1} Z. Wu, P. Li, Y. Zhang, H. Guo, Y. Gu, \textit{Multicharged
vortex induced in fractional Schrödinger equation with competing nonlocal
nonlinearities}, J. Optics \textbf{21} (2019) 105602

\bibitem{vort3} L. Dong, C. Huang, \textit{Vortex solitons in fractional
systems with partially parity-time-symmetric azimuthal potentials},
Nonlinear Dyn. \textbf{98} (2019) 1019--1028

\bibitem{vort4} Q. Wang, G. Liang, \textit{Vortex and cluster solitons
in nonlocal nonlinear fractional Schrödinger equation}, J. Optics
\textbf{22} (2020) 055501

\bibitem{vort5} P. Li, B.A. Malomed, D. Mihalache, \textit{Metastable
soliton necklaces supported by fractional diffraction and competing
nonlinearities}, Opt. Exp. \textbf{28} (2020) 34472-33488

\bibitem{Yingji1} S. He, B.A. Malomed, D. Mihalache, X. Peng, X.
Yu, Y. He, D. Den, \textit{Propagation dynamics of abruptly autofocusing
circular Airy Gaussian vortex beams in the fractional Schrödinger
equation}, Chaos, Solitons \& Fractals \textbf{142} (2021) 110470

\bibitem{Yingji2} S. He, B.A. Malomed, D. Mihalache, X. Peng, Y.
He, D. Deng, \textit{Propagation dynamics of radially polarized symmetric
Airy beams in the fractional Schrödinger equation}, Phys. Lett. A
\textbf{404} (2021) 127403

\bibitem{Zhong_PD23} M. Zhong, \textit{Two-dimensional fractional
PPT-symmetric cubic-quintic NLS equation: Double-loop symmetry breaking
bifurcations, ghost states and dynamics}, Physica D \textbf{448} (2023)
133727

\bibitem{Zhong_CHAOS23} M. Zhong, L. Wang, P. Li, Z. Yan, \textit{Spontaneous
symmetry breaking and ghost states supported by the fractional PT-symmetric
saturable nonlinear Schrödinger equation}, Chaos 33 (2023) 013106



\end{thebibliography}
\end{document}